\documentclass[10pt]{article}
 \UseRawInputEncoding 
\usepackage[table]{xcolor}
\usepackage{amsmath}
\usepackage{float}
\usepackage{graphicx}
\usepackage{tabularx}
\usepackage[letterpaper, top = 1.0in, bottom = 1.0in, right=0.75in, left = 0.75in]{geometry}
\usepackage{amssymb} 
\usepackage{titling}
\usepackage{titlesec}
\usepackage{times}
\usepackage{ragged2e}

\usepackage{blindtext}
\usepackage{graphics} 
\usepackage{epsfig} 
\usepackage{bm}
\usepackage{caption}
\usepackage{subcaption}
\usepackage[noadjust]{cite}
\usepackage[labelformat=simple]{subcaption}

\usepackage{dsfont}
\usepackage{relsize}
\usepackage{verbatim}
\usepackage{array}
\usepackage{hyperref}

\pretitle{ \begin{center} \Huge}

\posttitle{\end{center}}
\preauthor{\begin {center} \normalsize}

\postauthor{\end{center}}
\date{}
\titleformat{\section}{\normalsize \bfseries \scshape}{\thesection}{1em}{}
\titleformat{\subsection}{\normalsize \bfseries}{\thesubsection}{1em}{}
\title{\Huge Ford Highway Driving RTK Dataset: \\ 30,000 km of North American Highways }

\author{Sarah E. Houts\textsuperscript{1}, Nahid Pervez\textsuperscript{2}, Umair Ibrahim\textsuperscript{1}, Gaurav Pandey\textsuperscript{2}, Tyler G. R. Reid\textsuperscript{3}\\
\textsuperscript{1}~\textit{Ford AV, LLC}, 
\textsuperscript{2}~\textit{Ford Motor Company}, 
\textsuperscript{3}~\textit{Xona Space Systems} \\}

\begin{document}
\maketitle
\thispagestyle{empty}

\section*{BIOGRAPHIES}
\noindent \textbf{Sarah E. Houts} works on mapping and localization technologies for Autonomous Vehicles as a research engineer at Ford Autonomous Vehicles, LLC. She has a B.S. in Aerospace Engineering from UC San Diego, and a M.S. and Ph.D. in Aeronautics and Astronautics from Stanford University. During her Ph.D. she worked on localization and path planning for underwater autonomous vehicles with MBARI (Monterey Bay Aquarium Research Institute).

~\\
\noindent \textbf{Nahid Pervez} is a Supervisor in Partial Autonomy at Ford Motor Company in the Controls and Automated Systems organization. He leads a research team focusing on embedded software development, controls, and AI/ML for highly automated driving systems. Prior to this role Nahid worked in Powertrain Controls and Calibration group at Ford. Nahid’s graduate studies was on the Lake Michigan Offshore Wind Feasibility Assessment project with the U.S. Department of Energy. He completed his undergraduate degree from Bangladesh University of Engineering and Technology.

~\\
\noindent \textbf{Umair Ibrahim} works on Wireless Connectivity Research at Ford Autonomous Vehicle, LLC and is developing DSRC and Cellular-based V2X safety communications applications. Previously, Umair worked as Ford Technical lead Engineer with the Federal Highway Administration and Crash Avoidance Metrics Partnership (CAMP) to develop V2I based safety applications for deployment in various regions of the U.S. Umair is currently pursuing a Ph.D. in ECE from UC Santa Cruz, and holds M.S. EE from University of Minnesota ('13), and B.S. in Electronics Eng. ('07) from Pakistan.

~\\
\noindent \textbf{Gaurav Pandey} is a Technical Leader of autonomous vehicles research at Ford Autonomous Vehicles LLC. He leads a team working on developing algorithms for smart vehicles in a smart world. Prior to Ford, Dr. Pandey was an Assistant Professor at the Electrical Engineering department of Indian Institute of Technology (IIT) Kanpur in India. At IIT Kanpur he was part of two research groups (i) Control and Automation, (ii) Signal Processing and Communication. His current research focus is on visual perception for autonomous vehicles and mobile robots using tools from computer vision, machine learning and information theory. He did his B-Tech from IIT Roorkee ('06) and completed his Ph.D. from University of Michigan, Ann Arbor ('13).

~\\
\noindent \textbf{Tyler G. R. Reid} is a co-founder and CTO of Xona Space Systems, a start-up focused on GNSS augmentation from Low-Earth Orbit. Tyler previously worked as a Research Engineer at Ford Motor Company in localization and mapping for self-driving cars. He was also a Software Engineer at Google and a lecturer at Stanford University where co-taught the course on GPS. Tyler received his Ph.D. (’17) and M.Sc. (’12) in Aeronautics and Astronautics from Stanford where he worked in the GPS Research Lab and his B.Eng in Mechanical Engineering from McGill. He is a recipient of the RTCA Jackson Award.

\section*{ABSTRACT}
 There is a growing need for vehicle positioning information to support Advanced Driver Assistance Systems (ADAS), Connectivity (V2X), and Autonomous Driving (AD) features. 
 These range from a need for road determination ($<$5 meters), lane determination ($<$1.5 meters), and determining where the vehicle is within the lane ($<$0.3 meters). 
 This paper presents the Ford Highway Driving RTK (Ford-HDR) dataset. This dataset includes nearly 30,000 km of data collected primarily on North American highways during a driving campaign designed to validate driver assistance features in 2018. This includes data from a representative automotive production GNSS used primarily for turn-by-turn navigation as well as an Inertial Navigation System (INS) which couples two survey-grade GNSS receivers with a tactical grade Inertial Measurement Unit (IMU) to act as ground truth. The latter utilized networked Real-Time Kinematic (RTK) GNSS corrections delivered over a cellular modem in real-time. This dataset is being released into the public domain to spark further research in the community.


\section{INTRODUCTION} \label{sec:intro}


Autonomous vehicles and a growing number of increasingly automated driver assistance features push requirements for progressively increasing positioning accuracy~\cite{Reid2019c}. Today, Global Navigation Satellite Systems (GNSS) are used to provide position information as a driver navigational aid. This provides an attractive solution, as it offers global positioning using relatively low-cost hardware with lightweight computational load. In recent years, accuracy and robustness have increased, thanks to the availability of substantially more GNSS satellites, multiple civil frequencies such as L5, multi-frequency capable mass market receivers, and continental-scale coverage of corrections services like networked Real-Time Kinematic (RTK),  Precise Point Positioning (PPP), and other model based approaches such as PPP-RTK~\cite{joubert2020developments}.

One of the challenges facing adoption of RTK and other precision GNSS solutions in next-generation automotive systems is understanding the environment that vehicles will be operating in, as this could potentially be used as a core component of a safety critical system. General Motor's (GM) Super Cruise is an example use of GNSS as a core input to the feature activation criteria, only allowing the feature to be active on divided highways~\cite{Hay2018a}. In order to address the integrity of such a system, the GNSS conditions on roads in terms of service denials must be understood. Some of the factors that affect the performance of GNSS and RTK use on highways include obstructions (e.g. overpasses and overhead signs), Radio Frequency (RF) interference (unintentional or otherwise), and reliability of cellular connectivity to receive the corrections. Characterizing these factors requires real-world driving data that is geographically diverse covering rural, urban, and highway driving over a large number of kilometers. 

Here, we present the Ford Highway Driving RTK (HDR) dataset, an open source dataset intended for the navigation and self-driving research community to help answer these questions. This data  was collected as part of a larger campaign used to develop and validate other vehicle features. While this experiment was not designed with GNSS analysis in mind, the data still allows for insights into the current highway environment for GNSS performance. An analysis of this data was presented previously by Ford in 2019~\cite{Reid2019a}.

Such a large scale GNSS dataset has not been released publicly before. There are a few detailed datasets specifically for GNSS applications, but they are available only at a much smaller scale. For example, in 2019, Humphreys et al. released a 2-hour GNSS dataset of urban driving in Austin, Texas~\cite{Humphreys2019}. There are a number of Autonomous Driving datasets available, but they generally are focused on components of Autonomous Driving such as perception and do not have much in the way of GNSS-specific data, often providing post-processed GNSS + IMU data as ground truth only. These include Argoverse~\cite{Chang}, Apollo~\cite{Huang2018}, KITTI~\cite{Geiger2013}, the Canadian Adverse Driving Conditions (CADC)~\cite{Pitropov2020}, and the Ford multi-agent seasonal~\cite{Agarwal2020} open datasets . The authors hope to further the trend of the industry to share more data and enable other researchers to gain additional insights beyond the basics of accuracy, availability, and continuity. 

This paper provides a detailed description of the data collection, the dataset itself, and provides some example analyses. We close with some open questions to the research community that this dataset can either help answer or that can inform future Ford open datasets or data collection campaigns.

\section{Data Collection}

\begin{figure}
    \centering
    \includegraphics[width=0.8\textwidth]{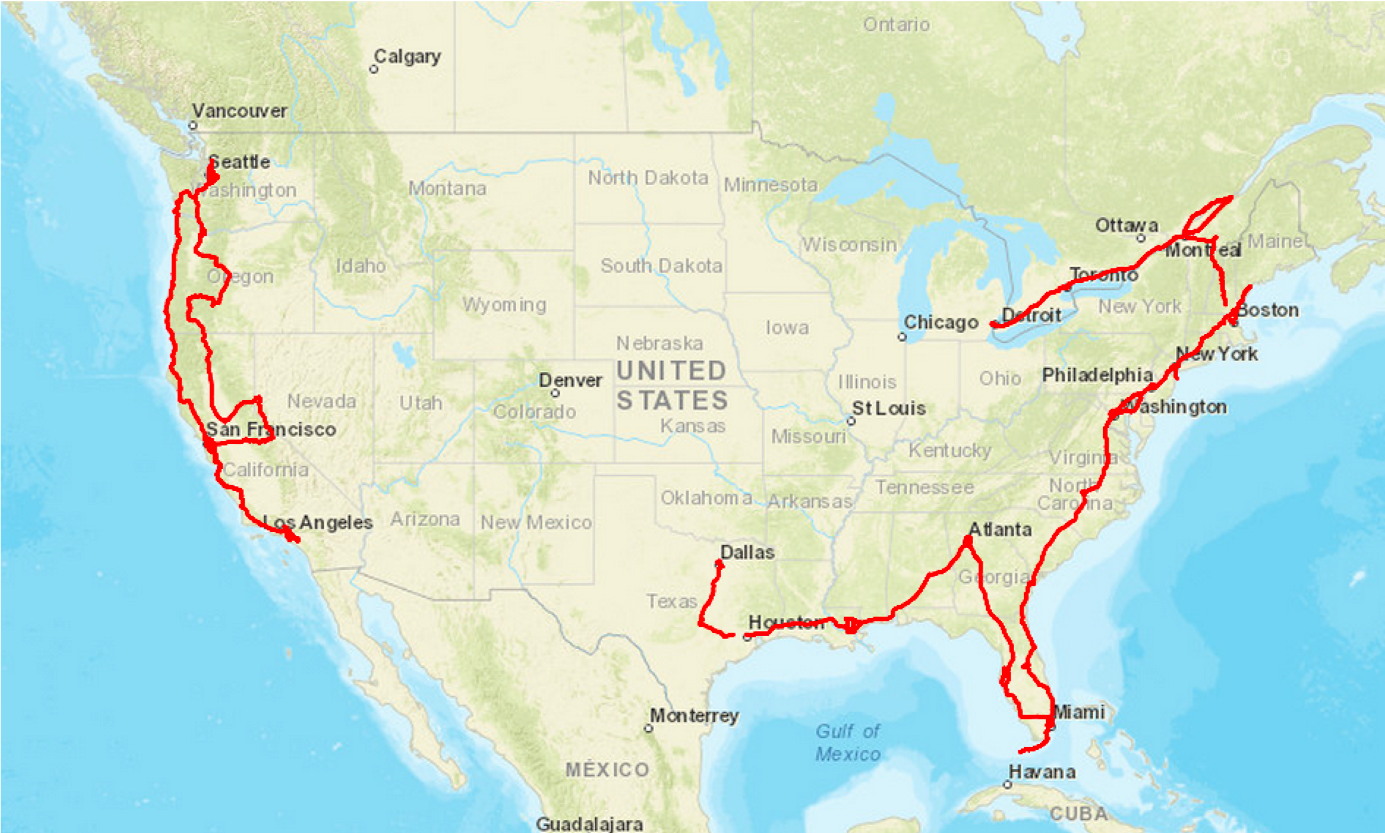}
    \caption{Map of the area covered by the Ford HDR dataset. Two vehicles were used, one on each coast.}
    \label{fig:area_driven}
\end{figure}

\begin{figure}
    \centering
    \includegraphics[width=0.8\textwidth]{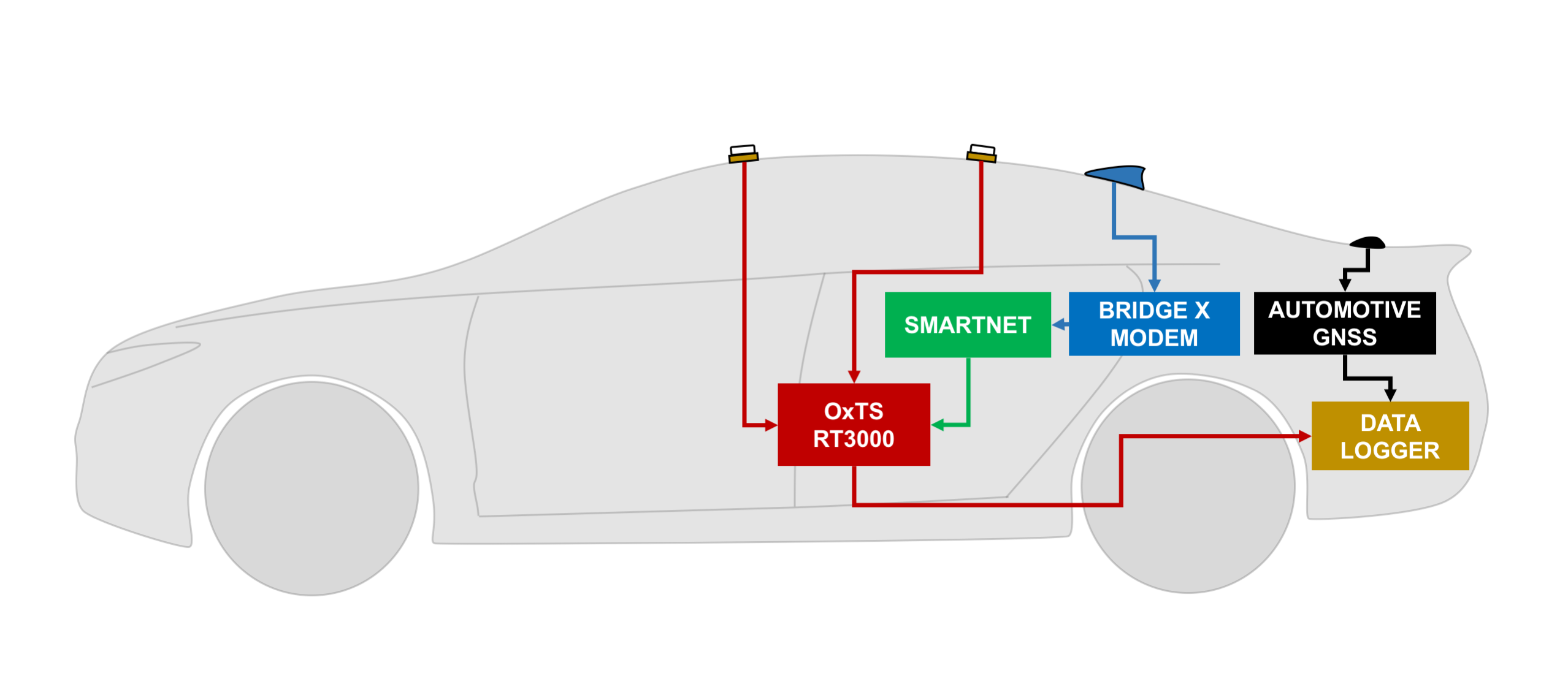}
    \caption{Schematic of the experimental vehicle setup. Antenna placement is representative, other equipment layout is spaced out for clarity and may not be representative of physical location on the vehicle.}
    \label{fig:vehicle_setup}
\end{figure}


The GNSS data presented here was part of a larger data collection campaign used for the development and validation of other vehicle features. This particular dataset was driven in mid-2018, primarily targeting the population centers of the east and west coasts of the U.S. and Canada on the route shown in Figure~\ref{fig:area_driven}. Two separate vehicles were used in the data collection, one on each coast. The experiment design and hardware setup were largely the same on both vehicles. A high-level overview of the data collection vehicle is shown in Figure~\ref{fig:vehicle_setup}. 

A total of 27,500 kilometers (17,088 miles) was driven, representing 355 hours ($\sim$15 full days) of data. This was mostly highway driving although some of the route did pass through major urban centers. It also includes a wide variety of highway environments, including challenging mountain roads. To put this number of miles in perspective, the U.S. National Highway System (NHS) consists of approximately 350,000 kilometers (220,000 miles) of road. Hence, the data shown here represents roughly 8\% of the NHS. 

Two separate GNSS systems were used in data collection, one representative of that used in production vehicles today and the other a survey-grade system to act as a reference or ground truth. The production-oriented GNSS was automotive grade, multi-constellation capable (GPS + GLONASS + Galileo), and utilized the L1 frequency only. The survey-grade system was an Oxford Technical Solutions (OxTS) RT3000. The RT3000 Inertial Navigation System (INS) combines two survey-grade GNSS receivers with an Inertial Measurement Unit (IMU) to provide the ability to coast through reasonable GNSS outages. The RT3000 units used in this experiment were GPS + GLONASS and capable of tracking both the L1 and L2 frequencies. The MEMS IMU is tactical grade with a gyro bias stability of 2 degrees / hour. The two GNSS receivers are used to constantly calibrate the attitude of the RT3000 IMU, primarily to provide heading.

Networked RTK corrections were also used in this experiment. Cellular connectivity was available on the vehicle and corrections were delivered to the RT3000 via an RTK Bridge-X modem from Intuicom Wireless Solutions. The networked RTK corrections were provided by Hexagon's SmartNet, where coverage was available along the entire data collection route.

Both systems provided high level GNSS information including UTC time, latitude, longitude, altitude, number of satellites tracked, and Dilution Of Precision (DOP). The production GNSS data was collected at 1 Hz and the RT3000 at 30 Hz. The RT3000 also output several other parameters related to the full inertial navigation solution including attitude (roll, pitch, heading), position mode (RTK fixed, RTK float, differential code, Standard Positioning Service (SPS), or no service), age of differential corrections, and the 1$\sigma$ uncertainty of the position solution. 


\section{Data Description}
\label{sec:DataDescription}

This dataset is available as set of \texttt{*.csv} files, shared on the Ford GNSS data website (currently at \url{http://ford-hdr-gnss.s3.amazonaws.com/ford-hdr-gnss.zip}). Each file consists of an array of data with a header line that contains the field names. 

\begin{figure}[h]
    \centering
    \includegraphics[width=0.5\textwidth]{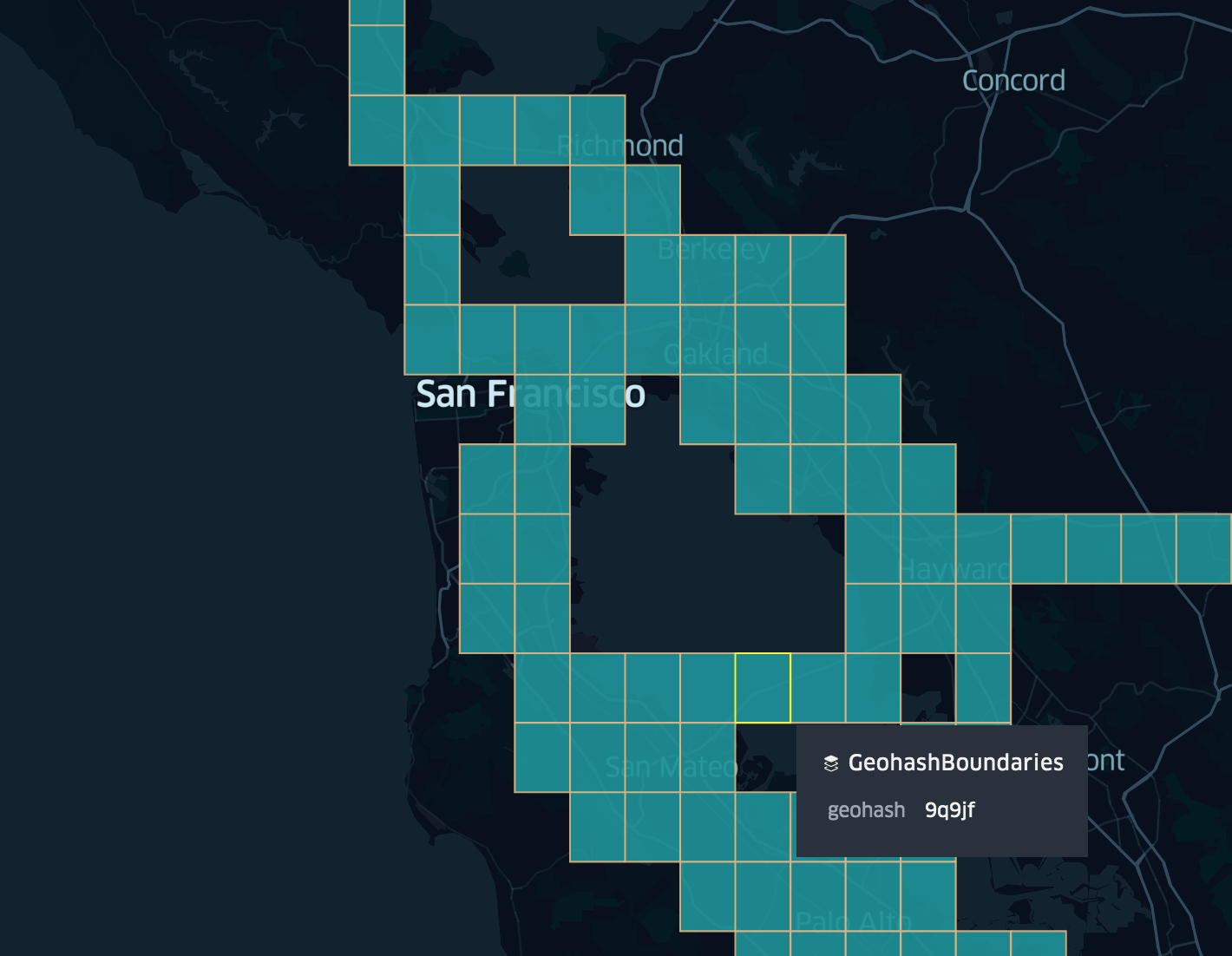}
    \caption{San Francsisco Bay Area geohash tiles. This depicts a small portion of the map available within the dataset showing all the tiles.}
    \label{fig:geohash}
\end{figure}

The first of these fields is \textit{geohash}, a unique identifier for each 4.8x4.8~km tile that uses a 5 character Morton code.  As an example, Figure~\ref{fig:geohash} shows the tiles for the San Francisco Bay Area in California. A complete map showing all the tiles is included with the dataset. 
The geohash tiles in Washington and part of Oregon begin with $c$. The rest of the tiles in Oregon, California, Texas, and Louisiana begin with $9$. The bulk of the east coast tiles begin with $d$, all the way through Detroit, with the exception of the tiles in Quebec, which begin with $f$.

As described above, there are two GNSS systems in the vehicle. The fields associated with the production system are preceded by a $P$, such as \textit{P\_GPS\_Heading}. Those from the OxTS RT3000 are preceded by an $R$, such as \textit{R\_AngleHeading}.

There are a few fields worth noting from the production GNSS system. The \textit{P\_GPS\_timestamp} field is formatted as yyyy-mm-dd hh:mm:ss. The field for the number of satellites, \textit{P\_GPS\_Sat\_num\_in\_view}, represents an expected number based on the constellation geometry, rather than the actual number being tracked by the system. \textit{P\_Gps\_B\_Fault} describes if there is a fault in the production GNSS system, with 1 representing a fault state, and 0 representing no fault. \textit{P\_Latitude} and \textit{P\_Longitude} are for the reference point on the vehicle, which is represented on the ground at the centroid of the vehicle.



To keep the dataset a manageable size, the data fields from the OxTS RT3000 INS were down-sampled to 1 Hz to match the data rate for the production system. The nearest timestamp from the original 30 Hz recording is shared in the \textit{R\_RT3k\_timestamp} field, which is formatted as yyyy-mm-dd hh:mm:ss.ddd. Many of the included fields from the RT3000 are related to the velocity and acceleration from the IMU, and are defined by the frames described in Section~\ref{sec:frames}. Additional information about the OxTS RT3000 outputs can be found in the manuals provided by the manufacturer~\cite{OxTS2015,OxTS2018}.

The only field (other than \textit{geohash}) that is not specifically for one system or the other is \textit{D\_Dist\_m}. This represents the distance between the Lat-Lon position of the two systems, reported in meters. This was calculated directly from the fields shown here, and thus does not account for the difference in reference points. The distance between the reference points can be calculated by estimating the lever arm between them~\cite{Reid2019a}.

\subsection{OxTS Frames}
\label{sec:frames}
As described in the OxTS user manual for RTv2 GNSS-aided inertial measurement systems \cite{OxTS2018}, three frames are used in this dataset. The first is the global North-East-Down (NED) navigation frame.
The OxTS navigation frame is attached to the vehicle but does not rotate with it. The Down axis is always aligned to the local gravity vector and North always points towards geodetic North. The fields in this frame are labeled with North, East, or Down, such as \textit{R\_VelNorth}.


The second frame is the OxTS local horizontal frame.
The OxTS horizontal frame is attached to the vehicle. The longitudinal and lateral axes remain parallel to a horizontal plane.  The fields in this frame are labeled with Lateral, Forward, or Down, such as \textit{R\_VelLateral}. Down for both the NED navigation frame and the horizontal frame point in the same direction. 


The final frame is the vehicle frame.
The OxTS vehicle frame is attached to the vehicle and rotates with it in all three axes. The X-axis remains parallel to the vehicle's heading, while the Y-axis points to the right and is perpendicular to the vehicle's vertical plan of symmetry. Fields in this frame are labeled with X, Y, or Z, such as \textit{R\_AccelX}. All of the OxTS frames have their origin at the center of the rear axle of the vehicle.


\section{Example Usage}

Already, some analysis has been performed on this dataset, and findings for on-road GNSS accuracy, availability, and continuity were presented previously in~\cite{Reid2019a}. Availability and continuity were broken down in terms of satellite visibility, satellite geometry, position type (RTK fixed, RTK float, or SPS), and RTK correction latency over the network. The results showed that current automotive solutions capable of meeting road determination requirements, but are not suitable for lane determination. A system using multi-frequency multi-constellation receivers with a correction service substantially increases capability.


\begin{figure}[h]
    \centering
    \includegraphics[width=.5\linewidth]{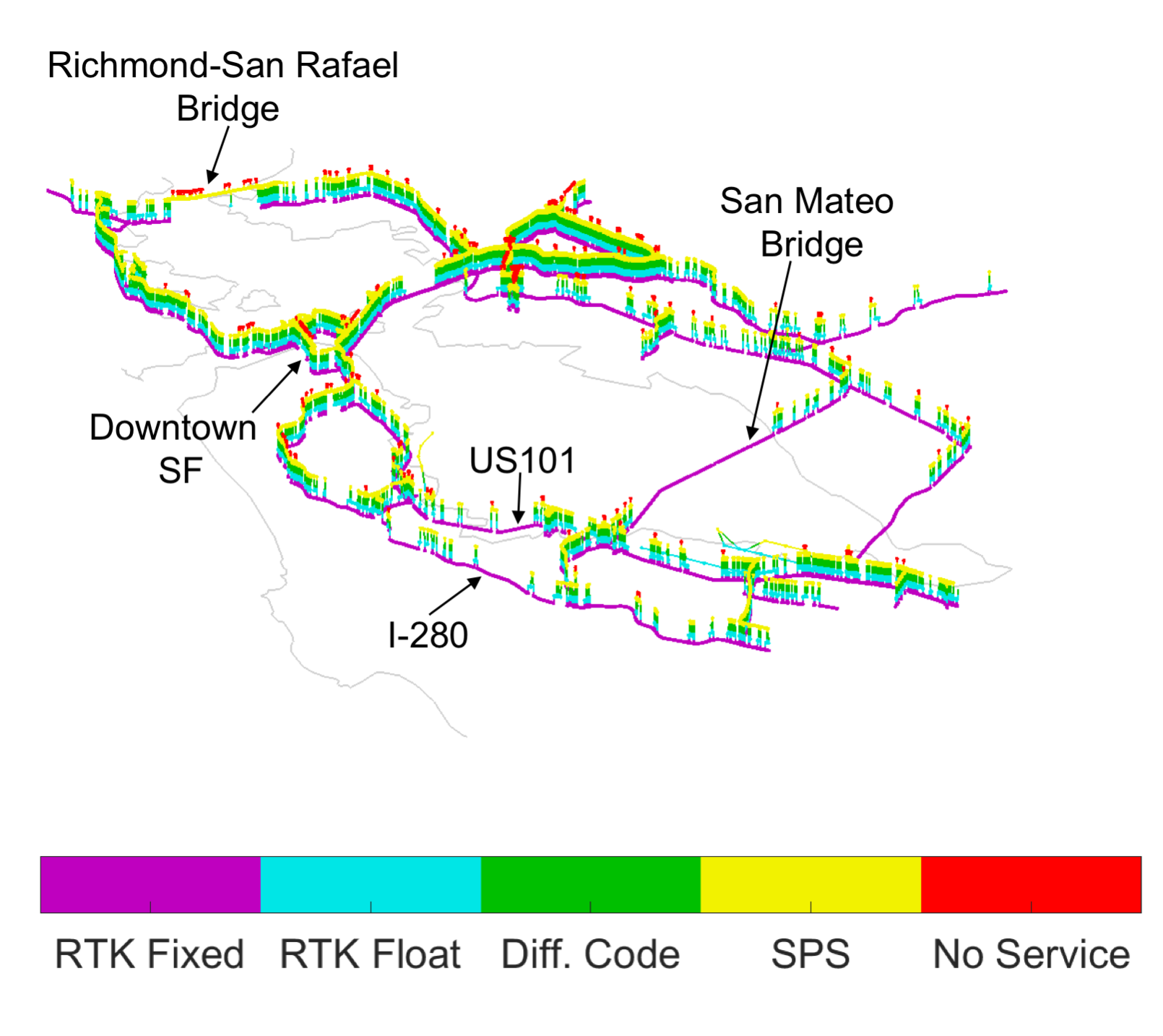}
    \caption{Bay Area Case Study. This example looks at the $R\_GpsPosMode$ field for the Bay Area.}
    \label{fig:bay-area-case-study}
\end{figure}

An example of the sorts of insights that can be gained from this data can be seen in Figure~\ref{fig:bay-area-case-study}. The $R\_GpsPosMode$ field is shown here in the San Francisco Bay Area. The difference between the sky view of the San Mateo Bridge and the Richmond-San Rafael Bridge (eastbound) can be seen in the resulting RTK mode. Along the San Mateo Bridge with an unobstructed sky, the RT3000 maintains RTK integer (fixed) positioning mode, while on the lower deck of the Richmond-San Rafael Bridge with a heavily obstructed sky view, it bounces between SPS and no service at all. Further insights could be gained by looking in closer detail at other locations where the system loses RTK positioning.


\section{Discussion and Conclusion}

The Ford Highway Driving RTK dataset represents a first in terms of large scale automotive GNSS data being publicly available. While this dataset has been evaluated for insights into GNSS performance on highways~\cite{Reid2019a}, the collected data was not intended for that use, and future data campaigns could be encouraged to make changes that enable additional GNSS-specific analysis. These could include usage of more recent GNSS hardware, particularly in terms of multi-constellation and multi-frequency elements; recording additional data types such as pseudorange or carrier phase; and extending into other road types, such as urban or rural settings. In the future, an equivalent dataset that includes urban centers could be used to compare expected and measured performance due to occlusions and multipath effects, providing validation for algorithms such as shadow matching. 

The hope of the authors is that other researchers will glean additional insights from this dataset. For example, combined with a map of overpasses and overhead signs, the locations where RTK-fixed mode is lost can be evaluated more extensively to understand if all such instances are due to environmental geometry or if there are other factors at play, such as other vehicles making use of Personal Privacy Devices (PPDs).

\section{Acknowledgements}
The authors would like to thank Ford Motor Company for supporting this work.

\bibliographystyle{ieeetran}
\bibliography{IEEEabrv,30k_dataset_bib}

\begin{thebibliography}{10}
\providecommand{\url}[1]{#1}
\csname url@rmstyle\endcsname
\providecommand{\newblock}{\relax}
\providecommand{\bibinfo}[2]{#2}
\providecommand\BIBentrySTDinterwordspacing{\spaceskip=0pt\relax}
\providecommand\BIBentryALTinterwordstretchfactor{4}
\providecommand\BIBentryALTinterwordspacing{\spaceskip=\fontdimen2\font plus
\BIBentryALTinterwordstretchfactor\fontdimen3\font minus
  \fontdimen4\font\relax}
\providecommand\BIBforeignlanguage[2]{{%
\expandafter\ifx\csname l@#1\endcsname\relax
\typeout{** WARNING: IEEEtran.bst: No hyphenation pattern has been}%
\typeout{** loaded for the language `#1'. Using the pattern for}%
\typeout{** the default language instead.}%
\else
\language=\csname l@#1\endcsname
\fi
#2}}

\bibitem{Reid2019c}
\BIBentryALTinterwordspacing
T.~G.~R. Reid, S.~E. Houts, R.~Cammarata, G.~Mills, S.~Agarwal, A.~Vora, and
  G.~Pandey, ``{Localization Requirements for Autonomous Vehicles},'' \emph{SAE
  International Journal of Connected and Automated Vehicles}, vol.~2, no.~3,
  pp. 173--190, sep 2019. [Online]. Available:
  \url{https://www.sae.org/content/12-02-03-0012/}
\BIBentrySTDinterwordspacing

\bibitem{joubert2020developments}
\BIBentryALTinterwordspacing
N.~Joubert, T.~G. Reid, and F.~Noble, ``Developments in modern gnss and its
  impact on autonomous vehicle architectures,'' in \emph{The 31st IEEE
  Intelligent Vehicle Symposium (IV)}, 2020. [Online]. Available:
  \url{https://arxiv.org/abs/2002.00339}
\BIBentrySTDinterwordspacing

\bibitem{Hay2018a}
C.~Hay, ``{Use of Precise Point Positioning for Cadillac Super Cruise},'' in
  \emph{Munich Satellite Navigation Summit}, Munich, 2018.

\bibitem{Reid2019a}
\BIBentryALTinterwordspacing
T.~G.~R. Reid, N.~Pervez, U.~Ibrahim, S.~E. Houts, G.~Pandey, N.~K.~R. Alla,
  and A.~Hsia, ``{Standalone and RTK GNSS on 30,000 km of North American
  Highways},'' in \emph{The 32nd International Technical Meeting of the
  Satellite Division of The Institute of Navigation (ION GNSS+ 2019)}, Miami,
  FL, sep 2019. [Online]. Available: \url{http://arxiv.org/abs/1906.08180}
\BIBentrySTDinterwordspacing

\bibitem{Humphreys2019}
\BIBentryALTinterwordspacing
T.~E. Humphreys, M.~J. Murrian, and L.~Narula, ``{Deep urban unaided precise
  Global Navigation Satellite System vehicle positioning},'' jun 2019.
  [Online]. Available: \url{http://arxiv.org/abs/1906.09539}
\BIBentrySTDinterwordspacing

\bibitem{Chang}
\BIBentryALTinterwordspacing
M.-F. Chang, J.~Lambert, P.~Sangkloy, J.~Singh, S.~B.~Ë. Ak, A.~Hartnett,
  D.~Wang, P.~Carr, S.~Lucey, D.~Ramanan, J.~Hays, and A.~Ai, ``{Argoverse: 3D
  Tracking and Forecasting with Rich Maps},'' Tech. Rep. [Online]. Available:
  \url{www.argoverse.org.}
\BIBentrySTDinterwordspacing

\bibitem{Huang2018}
\BIBentryALTinterwordspacing
X.~Huang, X.~Cheng, Q.~Geng, B.~Cao, D.~Zhou, P.~Wang, Y.~Lin, and R.~Yang,
  ``{The ApolloScape Dataset for Autonomous Driving},'' mar 2018. [Online].
  Available: \url{http://arxiv.org/abs/1803.06184}
\BIBentrySTDinterwordspacing

\bibitem{Geiger2013}
\BIBentryALTinterwordspacing
A.~Geiger, P.~Lenz, C.~Stiller, and R.~Urtasun, ``{Vision meets robotics: The
  KITTI dataset},'' \emph{The International Journal of Robotics Research},
  vol.~32, no.~11, pp. 1231--1237, sep 2013. [Online]. Available:
  \url{http://journals.sagepub.com/doi/10.1177/0278364913491297}
\BIBentrySTDinterwordspacing

\bibitem{Pitropov2020}
\BIBentryALTinterwordspacing
M.~Pitropov, D.~Garcia, J.~Rebello, M.~Smart, C.~Wang, K.~Czarnecki, and
  S.~Waslander, ``{Canadian Adverse Driving Conditions Dataset},'' jan 2020.
  [Online]. Available: \url{http://arxiv.org/abs/2001.10117}
\BIBentrySTDinterwordspacing

\bibitem{Agarwal2020}
\BIBentryALTinterwordspacing
S.~Agarwal, A.~Vora, G.~Pandey, W.~Williams, H.~Kourous, and J.~McBride,
  ``{Ford Multi-AV Seasonal Dataset},'' 2020. [Online]. Available:
  \url{https://arxiv.org/abs/2003.07969}
\BIBentrySTDinterwordspacing

\bibitem{OxTS2015}
\BIBentryALTinterwordspacing
OxTS, ``{NCOM Inertial and GNSS measurement systems - NCOM Manual},'' 2015.
  [Online]. Available:
  \url{https://www.oxts.com/wp-content/uploads/2017/06/ncomman.pdf}
\BIBentrySTDinterwordspacing

\bibitem{OxTS2018}
\BIBentryALTinterwordspacing
------, ``{RTv2 GNSS-aided inertial measurement systems User Manual},'' 2018.
  [Online]. Available: \url{https://www.oxts.com/app/uploads/2018/02/rtman.pdf}
\BIBentrySTDinterwordspacing

\end{thebibliography}


\begin{table}
\centering
{\rowcolors{2}{gray!90!white!10}{gray!70!blue!20}
\begin{tabular}{|p{.2\linewidth}|p{.08\linewidth}|p{.65\linewidth}|}
    \hline
    Column Header & Units & Description \\ 
    \hline\hline
    geohash & n/a & 5 character Morton code geohash. Represents 4.8 km x 4.8 km area \\ 
    \hline
    P\_GPS\_timestamp & n/a & Time for production GNSS formatted as yyyy-mm-dd hh:mm:ss \\
    \hline
    P\_GPS\_Hdop & n/a & Horizontal Dilution of Precision for Production GNSS \\
    \hline
    P\_GPS\_Pdop & n/a & Position dilution of precision for Production GNSS \\
    \hline
    P\_GPS\_Vdop & n/a & Vertical dilution of precision for Production GNSS\\
    \hline
    P\_GPS\_Heading & degrees & Heading in degrees (clockwise from North) for Production GNSS [0, 360)\\
    \hline
    P\_GPS\_Sat\_num\_in\_view & n/a & Number of satellites in view for Production GPS (expected number for clear sky, not current number being tracked) (GPS + GLONASS + Galileo)\\
    \hline
    P\_Gps\_B\_Fault & n/a & Fault in GNSS system - 0: No Fault, 1: Fault\\
    \hline
    R\_RT3k\_timestamp & n/a & Time for RT3000 formatted as yyyy-mm-dd hh:mm:ss.ddd. Provides the closest timestamp to production system, down-sampled from 30 Hz\\
    \hline
    R\_GpsNumSats &  & Number of satellites currently being tracked by RT3000 (GPS + GLONASS)\\
    \hline
    R\_GpsPosMode & n/a & Positioning mode - None=0, Search=1, Doppler=2, SPS=3, Differential=4, RTK float=5, RTK integer=6\\
    \hline
    R\_GpsVelMode & n/a & Velocity mode - None=0, Search=1, Doppler=2, SPS=3, Differential=4, RTK float=5, RTK integer=6\\
    \hline
    R\_GpsAttMode & n/a & GPS dual antenna attitude mode\\
    \hline
    R\_PosNorthStdev & m & Standard deviation of Position in North (m)\\
    \hline
    R\_PosEastStdev & m & Standard deviation of Position in East (m)\\
    \hline
    R\_PosDownStdev & m & Standard deviation of Position in Down (m)\\
    \hline
    R\_VelNorthStdev & m/s & Standard deviation of VelNorth (m/s)\\
    \hline
    R\_VelEastStdev & m/s & Standard deviation of VelEast (m/s)\\
    \hline
    R\_VelDownStdev & m/s & Standard deviation of VelDown (m/s)\\
    \hline
    R\_AngleHeadingStdev & degrees & Standard deviation of Heading angle (deg)\\
    \hline
    R\_AnglePitchStdev & degrees & Standard deviation of Pitch angle (deg)\\
    \hline
    R\_AngleRollStdev & degrees & Standard deviation of Roll angle (deg)\\
    \hline
    R\_PosAlt & m & Altitude (above MSL) in meters\\
    \hline
    R\_VelNorth & m/s & Velocity in the North direction (m/s)\\
    \hline
    R\_VelEast & m/s & Velocity in the East direction (m/s)\\
    \hline
    R\_VelDown & m/s & Velocity in the Down direction (m/s)\\
    \hline
    R\_Speed2D & m/s & Horizontal speed ($\sqrt{R\_VelNorth^2+R\_VelEast^2}$) (m/s)\\
    \hline
    R\_VelForward & m/s & Forward Velocity (horizontal frame) - positive forward on vehicle, rotates with heading but not pitch/roll\\
    \hline
    R\_VelLateral & m/s & Lateral Velocity (horizontal frame) - positive to right of vehicle, rotates with heading but not pitch/roll\\
    \hline
    R\_AccelX & m/s$^2$ & Acceleration in X direction (OxTS vehicle frame, points forward on the vehicle)\\
    \hline
    R\_AccelY & m/s$^2$ & Acceleration in Y direction (OxTS vehicle frame, points right on vehicle)\\
    \hline
    R\_AccelZ & m/s$^2$ & Acceleration in Z direction (OxTS vehicle frame, points down on vehicle)\\
    \hline
    R\_AccelForward & m/s$^2$ & Acceleration in Forward direction (OxTS horizontal frame - positive forward on vehicle, rotates with heading but not pitch/roll)\\
    \hline
    R\_AccelLateral & m/s$^2$ & Acceleration in Lateral direction (OxTS horizontal frame - positive to right vehicle, rotates with heading but not pitch/roll)\\
    \hline
    R\_AccelDown & m/s$^2$ & Acceleration in Down direction \\
    \hline
    R\_AccelSlip &  & Slip rate\\ 
    \hline
    
\end{tabular}
}
\label{table:part1}
\caption{Description of the Ford HDR dataset fields.}
\end{table}

\begin{table}
\centering
{\rowcolors{2}{gray!90!white!10}{gray!70!blue!20}
\begin{tabular}{|p{.2\linewidth}|p{.08\linewidth}|p{.65\linewidth}|}
    \hline
    Column Header & Units & Description \\ 
    \hline\hline
    R\_AngleHeading & degrees & Heading angle in degrees (clockwise from North) from RT3000 [0, 360)\\
    \hline
    R\_AnglePitch & degrees & Pitch angle in degrees from RT3000 [-90, 90]\\
    \hline
    R\_AngleRoll & degrees & Roll angle in degrees from RT3000 [-180, 180]\\
    \hline
    R\_AngRateX & deg/s & Angular rate about the X axis (OxTS vehicle frame, points forward on vehicle)\\
    \hline
    R\_AngRateY & deg/s & Angular rate about the Y axis (OxTS vehicle frame, points right on vehicle)\\
    \hline
    R\_AngRateZ & deg/s & Angular rate about the Z axis (OxTS vehicle frame, points down on vehicle)\\
    \hline
    R\_AngRateForward & deg/s & Angular rate about the Forward axis (OxTS horizontal frame - positive forward on vehicle, rotates with heading but not pitch/roll)\\
    \hline
    R\_AngRateLateral & deg/s & Angular rate about the Lateral axis (OxTS horizontal frame - positive to right vehicle, rotates with heading but not pitch/roll)\\
    \hline
    R\_AngRateDown & deg/s & Angular rate about the Down axis\\
    \hline
    R\_AngleTrack & degrees & Direction of motion in global frame in degrees (clockwise from North) (0, 360) For land vehicles, usually very close to heading angle\\
    \hline
    R\_AngleSlip & degrees & Slip angle in degrees - (equals Track minus Heading) Note: Slip angle will be close to 180deg when traveling backwards\\
    \hline
    R\_DistanceWithHold & m & Total distance traveled in meters, starting from when the RT3000 is turned on. Note: Distance with hold will not increase when the RT measures a speed less than 0.2 m/s \\
    \hline
    R\_AngAccelX & deg/s$^2$ & Angular acceleration about the X axis (OxTS vehicle frame, points forward on vehicle)\\
    \hline
    R\_AngAccelY & deg/s$^2$ & Angular acceleration about the Y axis (OxTS vehicle frame, points right on vehicle)\\
    \hline
    R\_AngAccelZ & deg/s$^2$ & Angular acceleration about the Z axis (OxTS vehicle frame, points down on vehicle)\\
    \hline
    R\_AngAccelForward & deg/s$^2$ & Angular acceleration about the Forward axis (OxTS horizontal frame - positive forward on vehicle, rotates with heading but not pitch/roll)\\
    \hline
    R\_AngAccelLateral & deg/s$^2$ & Angular acceleration about the Lateral axis (OxTS horizontal frame - positive to right vehicle, rotates with heading but not pitch/roll)\\
    \hline
    R\_AngAccelDown & deg/s$^2$ & Angular acceleration about the Down axis\\
    \hline
    D\_Dist\_m & m & Distance between Production and RT position in meters\\ 
    \hline
    P\_Latitude & degrees & Latitude from Production GNSS in decimal degrees\\
    \hline
    P\_Longitude & degrees & Longitude from Production GNSS in decimal degrees\\
    \hline
    R\_Latitude & degrees & Latitude from RT3000 in decimal degrees\\
    \hline
    R\_Longitude & degrees & Longitude from RT3000 in decimal degrees\\
    \hline
\end{tabular}
}
\label{table:part2}
\caption{Description of the Ford HDR dataset fields, continued.}
\end{table}

\end{document}